\documentclass{aa}
\usepackage{graphicx}
\usepackage{tabularx}
\usepackage{rotating}
\usepackage{txfonts}
\usepackage{comment}
\usepackage{lscape}
\usepackage{threeparttable}
\usepackage{xcolor}
\usepackage{subfigure}
\usepackage{caption}
\usepackage{float}
\usepackage[switch]{lineno}
\usepackage{ulem}

\newcommand{\hi}{H\,\textsc{i}}
\newcommand{\hii}{H\,\textsc{i} 21cm}
\newcommand{\apx}{$\sim$}

\newcommand{\target}{4C\,31.04}
\newcommand{\eg}[1]{\citep[e.g.][]{#1}}
\newcommand{\kmps}{km~s$^{-1}$}
\newcommand{\p}[1]{$^{-#1}$}
\newcommand{\pp}[1]{$^{#1}$}

\newcommand{\beq}{\begin{equation}}
\newcommand{\eeq}{\end{equation}}

\newcommand{\tspin}{{T$_{\rm spin}$}}

\newcommand{\Msun}{M$_\odot$}

\newcommand{\NHI}{N$_{\rm HI}$}
\newcommand{\mjypb}{mJy~beam$^{-1}$}

\begin{document}

   \title{Turbulent circumnuclear disc and cold gas outflow in the \\
   newborn radio source \target}

   \author{Suma Murthy\inst{1},
          Raffaella Morganti\inst{2,3},
          Tom Oosterloo\inst{2,3},
          Robert Schulz\inst{2,4},
          Zsolt Paragi\inst{1}
         }

   \institute{Joint Institute for VLBI ERIC, Oude Hoogeveensedijk 4, 7991 PD Dwingeloo, The Netherlands. \\
                \email{murthy@jive.eu}
        \and
             ASTRON, The Netherlands Institute for Radio Astronomy, Oude Hoogeveensedijk 4, 7991 PD Dwingeloo, The Netherlands.
        \and
            Kapteyn Astronomical Institute, University of Groningen, P.O. Box 800, 9700 AV Groningen, The Netherlands.
        \and
            Leiden University, Snellius, Niels Bohrweg 1, 2333 CA, Leiden, The Netherlands.
}

   \date{Received 3 April 2024, accepted 6 May 2024}


 \abstract{We present deep kiloparsec- and parsec-scale neutral atomic hydrogen (\hi) absorption observations of a very young radio source ($\leq 5000$ years), \target, using the Westerbork Synthesis Radio Telescope (WSRT) and the Global Very Long Baseline Interferometry (VLBI) array. Using $z=0.0598$, derived from molecular gas observations, we detect, at both kpc and pc scales, a broad absorption feature (FWZI $= 360$ \kmps) centred at the systemic velocity, and  narrow absorption (FWZI $= 6.6$ \kmps) redshifted by 220 \kmps, both previously observed. Additionally, we detect a new blueshifted, broad, shallow absorption wing. At pc scales, the broad absorption at the systemic velocity is detected across the entire radio source while the shallow wing is only seen against part of the eastern lobe. The gas has higher \hi\ column density along the eastern lobe than along the western one. The velocity dispersion of the gas is high ($\geq 40$ \kmps) along the entire radio continuum, and is highest ($\geq 60$ \kmps) in the region including the outflow and the radio hot spot. While we detect a velocity gradient along the western lobe and parts of the eastern lobe at PA \apx\ $5^\circ - 10^\circ$, most of the gas along the rest of the eastern lobe exhibits no signs of rotation. Earlier optical spectroscopy suggests that the optical AGN is very weak. We therefore conclude that the radio lobes of \target\ are expanding into a circumnuclear disc, partially disrupting it and making the gas highly turbulent. The distribution of gas is predominantly smooth at the spatial resolution of \apx 4 pc studied here. However, clumps of gas are also present, particularly along the eastern lobe. This lobe appears to be strongly interacting  with the clouds and driving an outflow \apx 35 pc from the radio core, with a mass-outflow rate of $0.3 \leq \dot{M} \leq 1.4$ \Msun\ year\p{1}. It is likely that this interaction has caused the eastern lobe to be rebrightened, giving the source an asymmetric morphology. We compare our observations with the predictions of a recent analytical model regarding the survival of atomic gas clouds in radio-jet-driven outflows and find that the existence of a subkpc-scale outflow in this case could imply inefficient mixing of the cold gas with the hot medium and high gas density, leading to very short cooling times. Overall, our study provides further evidence of the strong impact of radio jets on the cold interstellar medium (ISM) in the early stages of their evolution and supports the predictions of numerical simulations regarding jet--ISM interactions and the nature of the circumnuclear gas into which the jets expand. }


\titlerunning{Jet--ISM interaction in 4C\,31.04}
\authorrunning{Murthy et al.}

\maketitle
\section{Introduction} \label{introduction}

A pressing question in extragalactic astronomy refers to the role of active galactic nuclei (AGN) in the evolution of their host galaxies. This can only be addressed by quantifying the impact of different types of AGN on their hosts and the evolution of this impact over the life-cycle of an active supermassive black hole. In the case of radio-loud AGN, where radio plasma is emitted in the form of collimated jets, evidence is slowly building up that the young radio jets entering the interstellar medium (ISM) significantly impact the surrounding gas \citep[e.g.][and references therein]{Holt08, Aditya18a, Morganti18, Glowacki17, Maccagni17, Murthy21, Shih14, Kukreti22}.

In order to better understand the contribution of radio AGN in shaping their host galaxies and the evolution of this impact with that of the radio source, it is essential to study radio AGN with different properties and in different evolutionary stages using various probes of the multi-phase ISM at different spatial scales. Although the best way to advance our understanding in this respect is with statistical studies of a large number of sources, these studies are immensely time consuming. Instead, detailed investigations of individual sources ---tracing the kinematics, physical condition, and location of the affected gas--- in combination with theoretical models of the jet--ISM interaction readily help in obtaining an initial overview of the impact radio AGN have on the surrounding medium. \citep[e.g.][]{Mukherjee18a, Audibert23, Murthy19, Murthy22b, Morganti13, Morganti15, Morganti21, Dasyra15, Cicone14, Alatalo11, Zovaro19}. While this has led to significant progress in understanding the impact of radio jets on the ISM, further studies are needed in order to cover a broader parameter space, for example in terms of morphology, radio luminosity, age, and size of the radio sources.

In this context, the 21 cm transition of neutral atomic hydrogen (\hi) detected in absorption is a useful probe of the gas in the vicinity of radio jets \citep[see][for a detailed review]{Morganti18}. Apart from the opacity of the gas itself, the detection of atomic hydrogen in absorption depends primarily on the strength and shape of the background radio continuum source. Thus, in the case of radio sources that are bright and extended at parsec (pc) scales, \hii\ absorption can be used to study jet--ISM interactions at very high spatial resolution using very long baseline interferometry (VLBI) techniques \citep[e.g.][]{Schulz18, Schulz21} 

Here we present new results from a detailed study of atomic gas in the central region of a well-known young radio source, \target, which is found to be interacting with the ambient ISM \citep{Zovaro19}. This is the first part of a broader study of the cold gas in this source, and a study of the molecular gas will be presented in a companion paper (Murthy et al.\ in prep).

\target\ is a young radio AGN hosted by the gas-rich galaxy MCG 5-4-18, making it an interesting source for studying the possible interaction between the radio jets and the ISM. It is one of the few sources for which the age has been estimated both from the expansion of the hot spots and a spectral index analysis \citep{Giroletti03}. This age estimation is in the range of 500 -- 5000 years, making it a genuinely young radio source.

\begin{figure*}
\centering
\includegraphics[height=6.7cm]{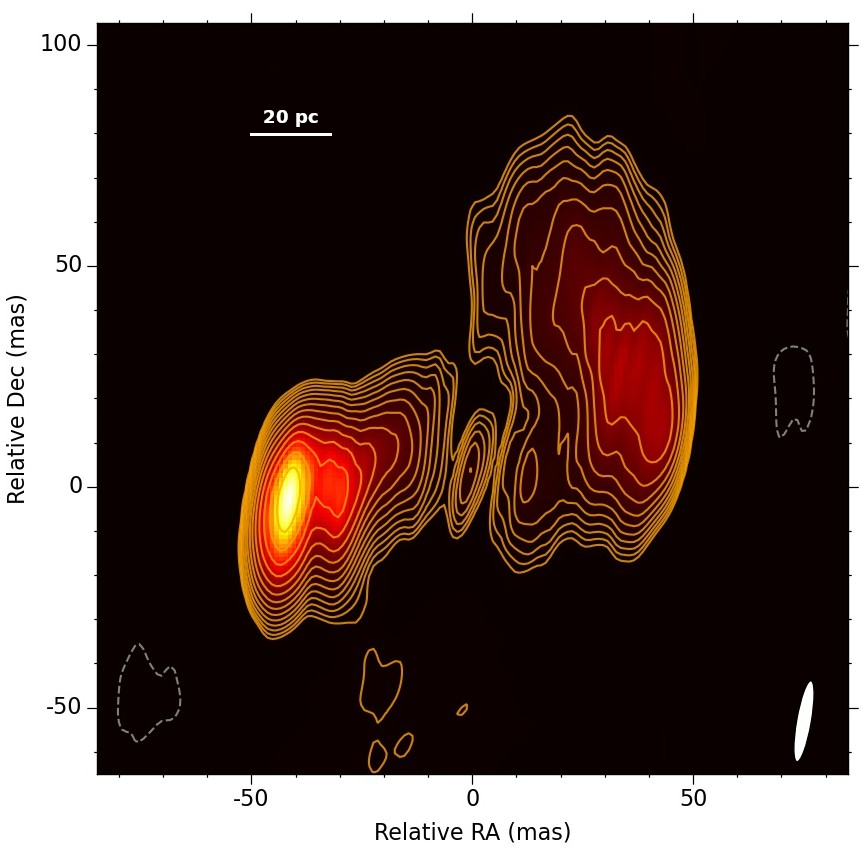} 
\includegraphics[height=7cm]{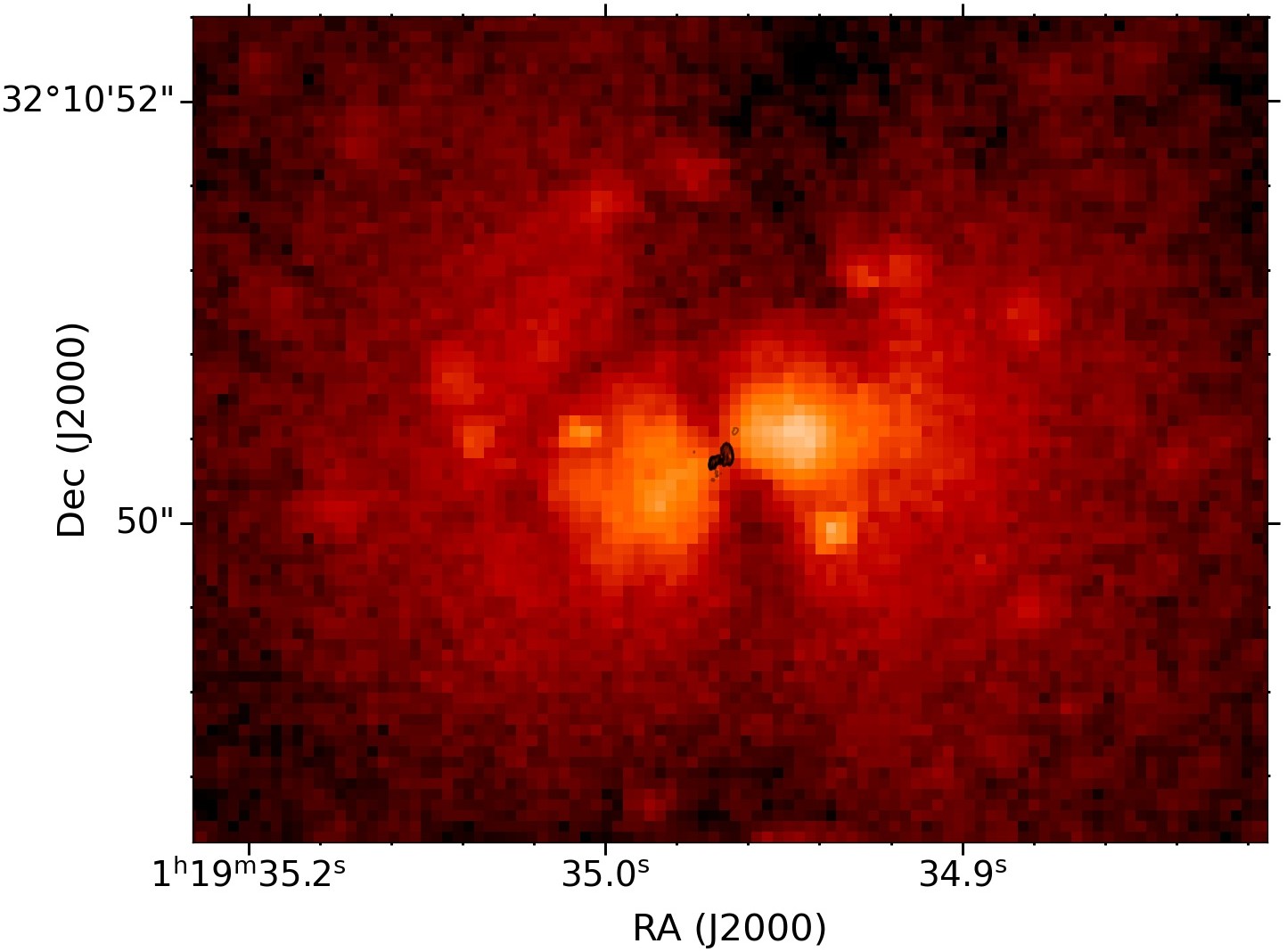} 
    \caption{\textbf{Left:} VLBI radio continuum map of \target. The contours start at 4$\sigma$ and increase by a factor of $\sqrt{2}$. The beam is shown in the bottom right corner of the image. The peak  brightness is 196 mJy beam\p{1}, measured at the hot spot along the eastern lobe. \textbf{Right:} HST WFPC2/PC zoomed-in image of the inner part of MCG 5-4-18 in colour overlaid with the radio source in the very centre surrounded by black contours. The HST image shows the warped dust lane perpendicular to the radio axis. The HST image is taken from the \textit{Mikulski} Archive for Space Telescopes (MAST).}
    \label{fig:Optical}
\end{figure*}

The radio source is \apx 140 pc in size\footnote{There is some uncertainty regarding the redshift of \target,\ with values in the literature ranging from $z=0.059$ \citep[e.g.][]{vanGorkom89} to $z=0.0602$ \citep[e.g.][]{Marcha96, Garcia-Burillo07}. Our molecular gas observations using NOEMA show a large-scale disc in CO(1-0) emission \citep{Morganti23a} from which we derive a redshift of $z=0.0598$, which we use in our analysis.} with two lobes and a core. It is reported to have a radio luminosity of  $1.14 \times 10^{25}$ W~Hz\p{1} at 5~GHz \citep{Giroletti03}, making it one of the weakest known compact symmetric objects \citep[CSOs;][]{Polatidis02}. Although at pc scales it consists of two lobes and a core, making it a CSO, the two lobes are of different morphology: the eastern lobe is the brighter of the two, includes a bright hot spot, and has a well-defined shape, while the western lobe is fainter overall, with a fainter hot spot, and has a relaxed, extended morphology (see Fig.~\ref{fig:Optical}, left).

The host galaxy, MCG 5-4-18, is  part of a small group, being the dominant member thereof. \textit{Hubble} Space Telescope (HST) observations \citep{Perlman01} show that both the nucleus and the outer regions of the galaxy are partly obscured by dust. In the nuclear region, cone-like features are present aligned with the radio lobes (Fig.~\ref{fig:Optical}, right), with the western side being redder than the eastern side. The origin of these features is not discussed by \citet{Perlman01} and they could simply be due to strong extinction from the molecular disc running north--south. Furthermore, there is another, elongated, warped obscuring feature perpendicular to the radio axis, extending 500 pc to the north and 1 kpc to the south.

A number of studies have shown that this galaxy is rich in cold gas. \citet{vanGorkom89} first detected a broad \hi\ absorption feature (FWHM $> 100$ \kmps) towards \target, which was later confirmed by \citet{Mirabel90}, who also detected an additional, narrow absorption feature (FWHM $\sim 6.6$ \kmps) redshifted with respect to the broad absorption, which they proposed was an infalling high-velocity cloud. \citet{Conway99} studied the \hi\ absorption towards this source at pc scales using the Very Long Baseline Array (VLBA) and proposed that the broad absorption arises from a disc with an inclination angle of $< 20^\circ$ , a thickness of 100 pc, and a radius of around 500 pc -- 1000 pc. \citet{Struve12}, analysing this further, only report a tentative velocity gradient perpendicular to the radio axis and conclude that the absorption arises from the atomic-gas counterpart of the molecular gas disc reported in the literature. 

Molecular gas has been detected from this source by various studies. \citet{Ocana-Flaquer10} report a molecular gas mass of $6.16 \times 10^8$ \Msun\ based on the detection of strong CO(1-0) emission and absorption with the IRAM 30m telescope. \citet{Garcia-Burillo07} used the Plateau de Bure interferometer to detect a 1.4 kpc molecular gas disc seen in HCO\pp{+} emission and absorption and suggested that the disc may not be dynamically relaxed owing to a jet--ISM interaction.

A study of warm molecular gas by \citet{Zovaro19} further supports the jet--ISM interaction scenario. These authors report the detection of shock-excited, warm molecular H$_2$ out to \apx 1 kpc, which they attribute to the far-reaching impact of the jet--ISM interaction, drawing inspiration from the numerical simulations that predict such an effect in the case of radio jets expanding in a clumpy ISM \citep[e.g.][]{Sutherland07, Bicknell18, Mukherjee16}. Thus, \target\ is particularly interesting for high-spatial-resolution studies of cold gas, where this jet--ISM interaction can be traced directly.

We observed \target\ with the wide band of the Westerbork Synthesis Radio Telescope (WSRT) in 2004 and detected a shallow, broad, blueshifted absorption feature in addition to the broad and the narrow features previously identified. This new feature ---an indicator of a fast outflow--- triggered more sensitive VLBI observations to localise the outflow and to study its properties and the possible impact of the radio jets on cold gas at pc scales.

We use a redshift of $z=0.0598$\footnote{At this redshift, 1$''$ corresponds to 1.13 kpc, assuming a flat Universe with H$_{0} = 70$ \kmps\ Mpc$^{-1}$, $\Omega_\Lambda = 0.7$, $\Omega_{\rm M} = 0.3$.} for \target\ based on our molecular gas observations, which will be presented in a companion paper (Murthy et al. in prep). We detail the observations and data reduction in Sect. 2, report our results in Sect. 3, discuss their implications for the jet--ISM interaction in Sect. 4, and finally conclude in Sect. 5.

\section{Observations and data reduction}
\label{sec:observations}

\subsection{WSRT observations}
\label{wsrtHI}
The WSRT observations of \target\ were carried out on April 11, 2004. The on-source time was five hours and the setup included a bandwidth of 20 MHz subdivided into 256 channels centred at 1341 MHz. The data reduction was carried out using Miriad \citep{Sault95} following standard procedures \eg{Morganti16}. Antennas C and D, giving the longest baselines in the array, had to be flagged while reducing the data. Although this limits the spatial resolution of the final images and cubes --- as \target\ is unresolved at arcsecond scales --- there is no spatial information lost as a consequence of the exclusion of those antennas.

We made the final continuum image and the cube using uniform weighting. The final continuum image has an RMS noise of 1.5 \mjypb. The final cube was made at a velocity resolution of 18 \kmps, with an RMS noise of 1.1 mJy beam\p{1} channel\p{1} and a beam size of 82$^{''} \times$ 14$^{''}$ with PA=4.9$^{\circ}$.

\subsection{Global VLBI observations}
\label{vlbiHI}

The phase-referenced Global VLBI observations were carried out on 24 May, 2019 with 17 stations of the EVN and VLBA arrays (Project code: GM045B). The two arrays observed this project for 10.3 hours each, with an overlap of three hours between the EVN and VLBA antennas. \target\ was observed for 12.86 hours in total with interleaved scans on the phase-reference calibrator, J0123+0344. We used 3C\,84 as a fringe finder and bandpass calibrator. The experiment was correlated into two data sets: a continuum pass with a total bandwidth of 64 MHz starting from 1315.75 MHz subdivided into four sub-bands with 32 spectral channels each, and a spectral-line pass of 16 MHz centred at 1339.75 MHz and subdivided into 512 spectral channels.

The data reduction was carried out using standard VLBI data reduction techniques \eg{Murthy19, Schulz18, Schulz21} in `classic' AIPS \citep[][]{Greisen03}. For both the continuum and line passes, we first applied the parallactic-angle and a priori gain corrections, and initial flagging using the calibration and flag tables obtained from the EVN pipeline and then corrected for the ionospheric dispersive delays. For the continuum pass data, we additionally corrected for the instrumental delays. We subsequently flagged bad data followed by a global fringe fit to correct for phase delay and rate, and then estimated the bandpass solutions. After applying these solutions to the target \textit{uv} data, we used only the line-free channels to iteratively improve the gain solutions via self-calibration involving imaging and phase-only self-calibration cycles with the solution interval for the phase-only self-calibration being progressively reduced from 30 minutes to 1 minute. We then applied a single round of amplitude and phase self-calibration. Next, we subtracted the continuum from the visibilities and flagged the bad data in the residuals, copied the flag table, and then repeated the self-calibration cycle mentioned above. We continued this until the continuum image showed no further improvement. The final continuum image presented in this paper was made with natural weighting using the continuum-pass data. In the case of the line pass, we subtracted the final continuum model from the visibilities and further subtracted a first-order polynomial from each visibility spectrum to remove any residual continuum emission. We note that due to the fact that the complex structure of the radio source limits accurate continuum subtraction, the noise on the spectra are not perfectly Gaussian. Finally, we deredshifted the \textit{uv} data using $z = 0.0598$. Next we imaged each channel using natural weighting to obtain the final spectral cube using the same restoring beam as the continuum image. 

The continuum image has an RMS noise of 330 $\mu$Jy~beam\p{1}, and the RMS noise on the cube is 460 $\mu$Jy~beam\p{1} per 6.98~\kmps channel. Both have a restoring beam of 12.68~mas~$\times$~3.18~mas with PA $=-13.25^{\circ}$. The details of observations, images, and cubes are listed in Table \ref{tab:obs_info}. 

The detected absorption is well resolved against the bright radio source. Hence, we used the Source Finding Application \citep[SoFiA;][]{Serra15, Westmeier21}\footnote{https://github.com/SoFiA-Admin/SoFiA.} to extract moment maps from only the broad absorption feature in order to further study the kinematics of the gas.
We extracted the moment maps using a mask to include all absorption beyond 4$\sigma$ in the line channels from the region co-spatial with the radio source. Table~\ref{tab:obs_info} lists the observation details as well as the image and cube parameters.

\begin{table*}
\caption{Observations and imaging parameters}
\centering
\begin{tabular}{ccccccccccc}
\hline\hline
Telescope & $\nu_{\rm{obs}}$ & on-source & BW  & $\Delta v$ & Weighting & beam-size & PA & RMS$_{\rm{cont}}$ & RMS$_{\rm{cube}}$ \\  & (MHz) & (hours) & (MHz) &(km s$^{-1}$) & & $''\times''$ & ($^\circ$) & ($\mu$Jy beam\p{1}) & (mJy beam\p{1})\\ (1) & (2) & (3) & (4) & (5) & (6) & (7) & (8) & (9) \\ \hline

WSRT  & 1341 & 5 &  20 & 18  & Uniform & 82 $\times$ 14 & 4.9 & 1500 & 1.1\\
G-VLBI$^*$ & 1339 & 12.9  &  16   & 7 & Natural & 0.0127 $\times$ 0.0032 & -13.25 & 330 & 0.46\\
 \hline
\end{tabular}
\begin{tablenotes}
\item The columns are: (1) Telescope used; (2)  central frequency of the band; (3)  on-source time; (4) bandwidth; (5) velocity resolution; (6) beam size; (7) beam position angle; (8)  RMS noise on the continuum map; and (9)  RMS noise on the spectral cube at a velocity resolution mentioned in (5).
\item $^*$Global VLBI observations. Participating stations: Effelsberg (Germany); Westerbork (the Netherlands); Jodrell-Bank (the United Kingdom); Onsala (Sweden); Torun (Poland); the eMERLIN stations: Cambridge, Darnhall, Pickmere (the United Kingdom); the VLBA stations: Hancock, North Liberty, Fort Davis, Los Almos, Kitt Peak, Pie Town, Owens Valley, Brewster, Mauna Kea (USA).
\end{tablenotes}
\label{tab:obs_info}
\end{table*}

\section{Results}\label{sec:results}

\subsection{WSRT at kpc scales}

\begin{figure}
    \includegraphics[width=\linewidth]{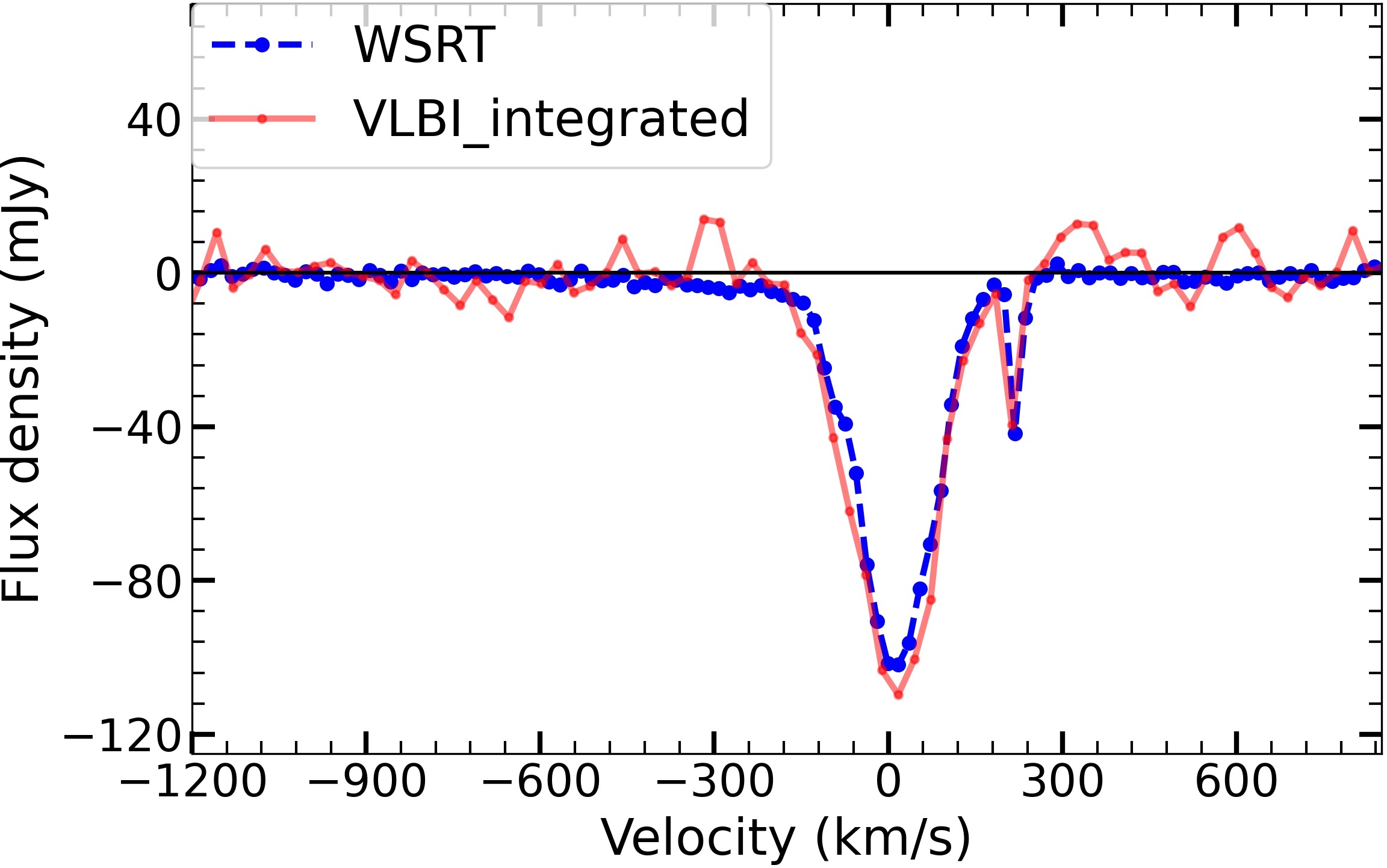}
    \caption{Comparison of the WSRT  and the global VLBI integrated \hi\ absorption profiles. The WSRT spectrum has a velocity resolution of 18 \kmps\ and an RMS noise of 1.1 mJy while the VLBI spectrum has a velocity resolution of 28 \kmps and an RMS noise of 10 mJy.}
    \label{fig:wsrt_evn_int}
\end{figure}

The radio continuum is unresolved at the kpc-scale resolution of WSRT and has a peak brightness of 2.67 Jy beam\p{1}, which agrees with the value reported by \citet{Mirabel90}. The \hi\ absorption profile is shown in Fig.~\ref{fig:wsrt_evn_int}. For $z~=~0.0598$, the profile includes the broad absorption centred at the systemic velocity  and the narrow redshifted absorption at 220~\kmps,  which are known from earlier studies. Additionally, we see a shallow, broad blueshifted wing not detected by earlier studies and extending up to $-400$~\kmps\ . The broad absorption profile and the shallow blueshifted wing combined have a full width at zero intensity (FWZI) of \apx 600 \kmps. We estimated the \hi\ column density using the standard relation: $N_{\hi} = 1.82 \times 10^{18}~(T_{spin}/c_f) \int \tau dv$~cm$^{-2}$, where we assume a gas spin temperature of \tspin\ $ = 100$~K, the typical temperature of the cold neutral medium \citep[see][]{Morganti18}, and the covering factor, c$_f = 1$. We estimated the total optical depth by integrating over the entire absorption profile, and using a peak brightness of 2.67 Jy beam\p{1}. We obtain $N_{\hi} =(1.04 \pm 0.09) \times 10^{21}$ cm\p{2} for the broad absorption centred at the systemic velocity. This is consistent with the earlier estimations in the literature \eg{Mirabel90}.

\begin{figure*}
\centering
    \includegraphics[width=8.75cm]{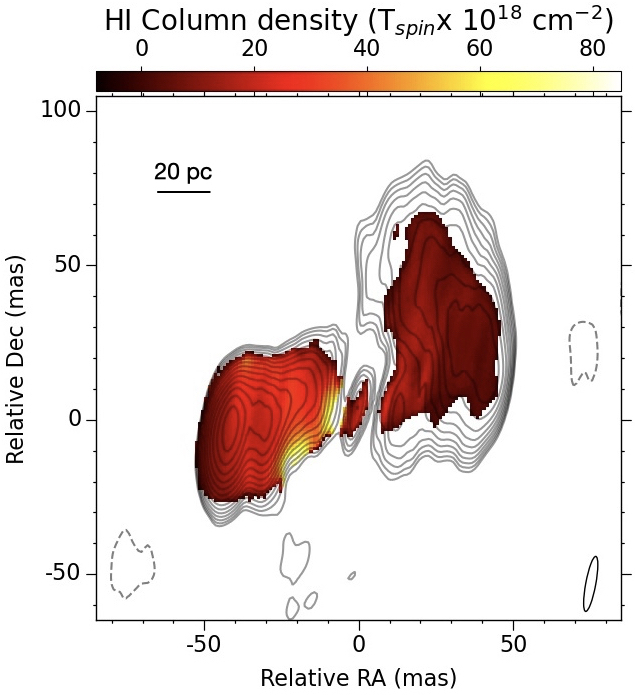}
    \includegraphics[width=9cm]{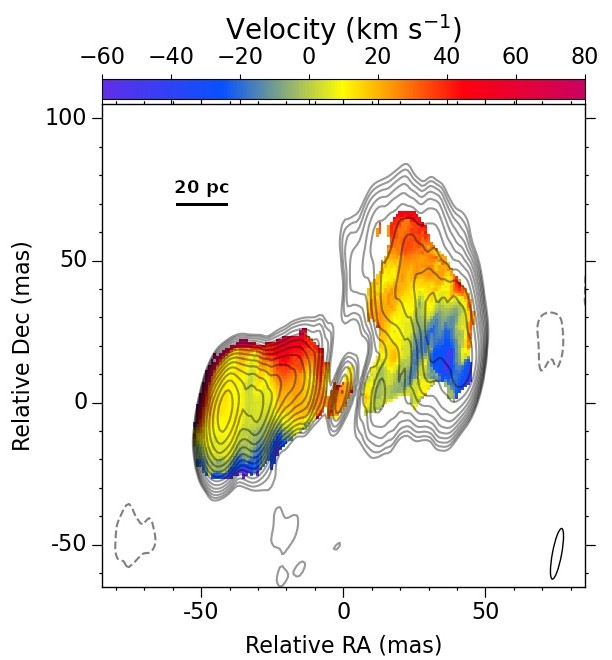}
    \caption{\textbf{Left:} \hi\ column density map of the gas seen in absorption in our VLBI data estimated assuming a gas spin temperature of 100 K. \textbf{Right:} Velocity field of the gas detected in absorption, shown in colour. There is a slight velocity gradient along the western lobe and also along the parts of the eastern lobe closer to the core in the north--south direction while such a gradient is entirely missing along the rest of the eastern lobe.  The contours in both images represent the radio continuum with the same contour levels as those in Fig.~\ref{fig:Optical}.}
    \label{fig:nhi_map}
    \label{fig:vel_field}
\end{figure*}

For the narrow absorption profile, we obtain \NHI $= (5.6 \pm 1.1) \times 10^{19}$ cm\p{2}. This too is consistent with earlier studies by \citet{Mirabel90} and \citet{Conway99} who concluded that the absorber is a foreground gas cloud far away from the nuclear region. As our focus is on the jet--ISM interaction, hereafter we only consider the broad absorption, which, as we discuss below, arises from the nuclear gas co-spatial with the radio source.

\begin{figure*}
\centering
    \includegraphics[height=11cm]{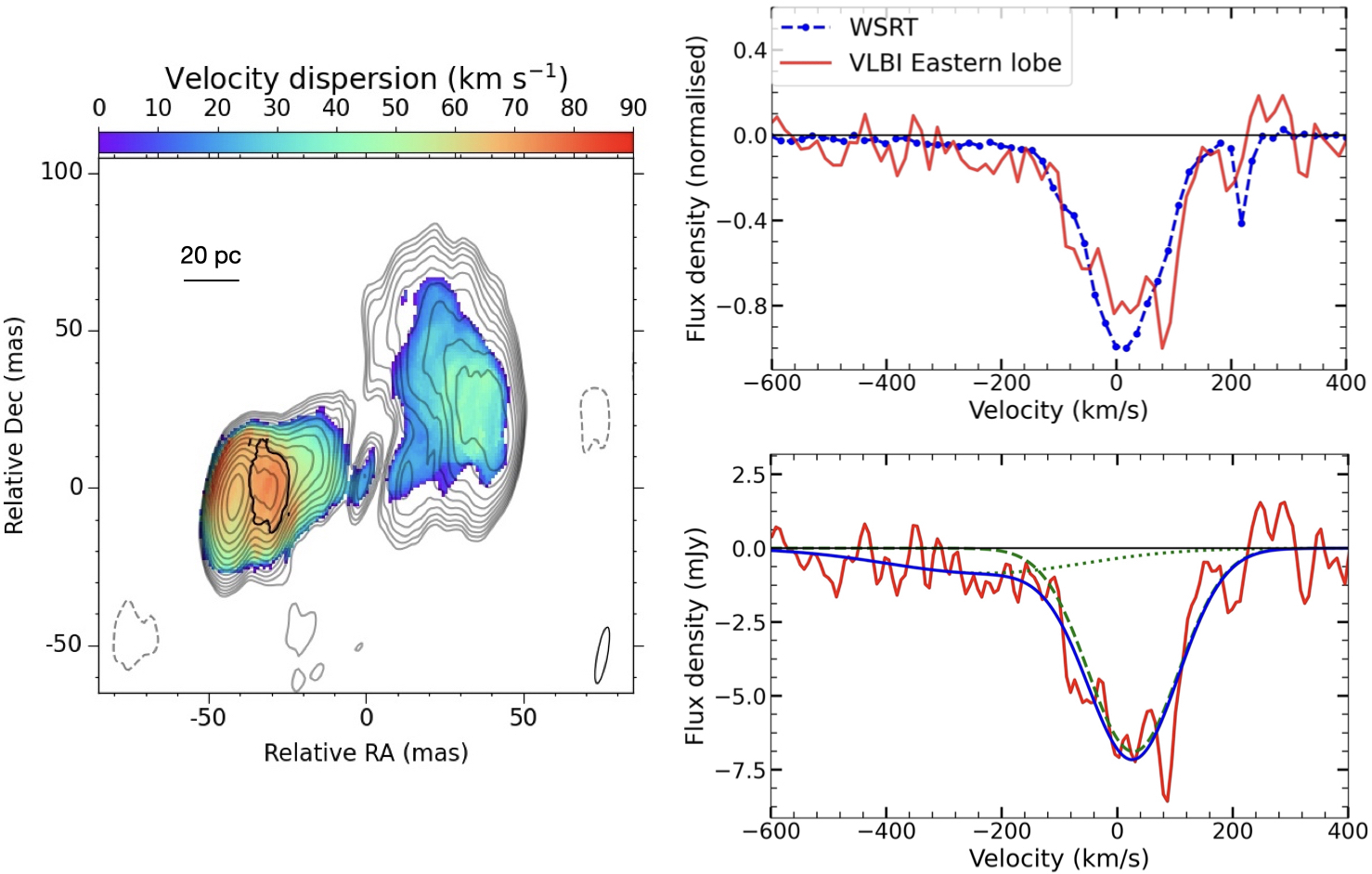}
    \caption{\textbf{Left:} Velocity dispersion map of the \hi\ seen in absorption with our VLBI data, shown in colour. The grey contours represent the radio continuum with the same contour levels as those in Fig.~\ref{fig:Optical}. The region where the shallow blueshifted wing is detected is surrounded with a black contour; this was ascertained by visual inspection. \textbf{Top right:} VLBI spectrum extracted from the region marked with the black contour in the velocity dispersion map --- where the blueshifted, shallow wing is detected --- compared with the WSRT spectrum. The spectra are normalised to emphasise the shallow wing in the VLBI spectrum .
    \textbf{Bottom right:} Gaussian fits to the deep and the shallow absorption features in the spectrum extracted from the region marked in black contour in the velocity dispersion map. The RMS noise on the VLBI spectrum is 0.7 mJy. See Sect.~\ref{sec:outflow} for details.}
    \label{fig:vel_disp_outflow}
    \label{fig:gauss_fits}
\end{figure*}

\subsection{Global VLBI at parsec scales}
\label{sec:vlbi_results}

The pc-scale VLBI continuum map is shown in the  left
panel of Fig.~\ref{fig:Optical}. The radio source has a projected size of \apx 140 pc with two lobes and a core, which is in agreement with earlier studies \eg{Conway96, Giroletti03}. Assuming an uncertainty on the flux density scale of 10\%, we measure an integrated flux density of $1.9 \pm 0.2$ Jy and a peak brightness of $196 \pm 20$ mJy beam\p{1}. \citet{Cotton95} report a higher total flux measurement of 2.46 Jy at  slightly lower spatial resolution (with a beam size of 17 mas $\times$ 10 mas, PA$=72^\circ$) with the VLBA at 1.8 GHz. \citet{Struve12} however report a significantly lower value of 1.16 Jy. As no variability is observed in \target\ \citep[][]{Giroletti03}, we conclude that the lower value of \citet{Struve12} is likely due to a combination of the lower sensitivity of their observations and calibration artefacts. The core has a total flux density of $13.5 \pm 1.4$ mJy. The total flux density within the eastern lobe is $1.0 \pm 0.1$ Jy and that within the western lobe is $900 \pm 90$ mJy. We detect both the redshifted narrow feature, and the broad feature centred at the systemic velocity against most of the radio source. 
The integrated \hi\ spectrum from our observations is shown in Fig.~\ref{fig:wsrt_evn_int}.  We measure FWZI $= 360$ \kmps\ for the absorption feature at the systemic velocity. We recover all the flux of the absorption detected with the WSRT. 

The left panel of Figure \ref{fig:nhi_map}  shows the \NHI\ map of the gas. The column density of the gas along the eastern lobe is higher compared to that along the western lobe with N$_{\rm H\textsc{I}, average}$ $= 2.5 \times 10^{21}$ cm\p{2}  for the former and  
N$_{\rm H\textsc{I}, average}$ = $7 \times 10^{20}$ cm\p{2} for the latter. At the location of the core, the column density is \NHI\ = $1.6 \times 10^{21}$ cm\p{2}. These estimations are for \tspin~$=~100$~K. \citet{Struve12}  reported a similar difference in the column density between the two lobes. Their values of \NHI\ for the core and the eastern lobe agree with our estimates, although these authors do not report an estimation for the western lobe. 

The velocity field of the gas is shown in the right
panel of Fig.\  \ref{fig:vel_field}. The gas along the western lobe and parts of the eastern lobe close to the core shows a velocity gradient along PA \apx $5^\circ - 10^\circ$, which is consistent with the molecular gas disc presented by \citet{Garcia-Burillo07} in terms of the sense of rotation and velocity gradient. On the other hand, the gas in the brightest part of the eastern lobe --- including the hot spot --- exhibits no significant gradient. We use these features to discuss the morphology of the absorbing gas in Sect.~\ref{sec:origin_broad_abs}.

The velocity dispersion of the absorbing gas is shown in the left panel of Fig. \ref{fig:vel_disp_outflow}. The gas seen against the eastern lobe exhibits a higher velocity dispersion compared to that along the western lobe. Interestingly, we detect the shallow, blueshifted absorption wing against only a part of the eastern lobe, \apx 31 mas from the core, corresponding to a projected distance of \apx 35 pc. The region from which this absorption arises is marked in the left panel of Fig. \ref{fig:vel_disp_outflow}. The width of this broad blueshifted wing is lower than that of the wing detected with the WSRT (see Sect.~\ref{sec:outflow}). This is due to the lower sensitivity of the VLBI \hi\ spectrum, which prevents us from detecting the faintest part of the blueshifted wing. We further note that, since the shallow absorption feature is detected over only a small region, it is not visible in the integrated spectrum (Fig. \ref{fig:wsrt_evn_int}). We fitted Gaussian components to the shallow and the deep absorption features in the VLBI spectrum extracted from the region containing the blueshifted wing. To quantify the significance of the shallow absorption feature, we used the Bayesian information criterion (BIC). We use $BIC = \chi^2 + k~ln(n),$ where $\chi^2$ is the chi-square statistic of the model, $n$ is the sample size, and $k$ is the number of model parameters. The two-component Gaussian model gives BIC$=-1.35,$ while for the single-component Gaussian model BIC$=7.57,$ with the absolute BIC difference between the two models being 8.9, which constitutes strong evidence in favour of the two-component Gaussian model \citep{Kass95}. The fits to the shallow and the deep absorption features are shown in the bottom-right panel of Fig.~\ref{fig:vel_disp_outflow}. We use this information in Sect.~\ref{sec:outflow}.

\section{Discussion}
\label{sec:discussion}

We detect multiple \hi\ absorption features in our observations: a narrow feature redshifted by 220 \kmps\ with respect to the systemic velocity, a deep, broad absorption centred at the systemic velocity for $z=0.0598$, and a shallow, broad, blueshifted absorption wing. The broad shallow wing is a new detection and the high spatial resolution achievable by the Global VLBI array further allows us to localise it. As mentioned above, the focus of this paper is on the gas giving rise to the broad, shallow absorption feature and the feature centred at the systemic velocity. We first discuss the morphology of the gas giving rise to the latter absorption feature in Sect. \ref{sec:origin_broad_abs} and then discuss the former in Sect.~\ref{sec:outflow}.

\subsection{A disturbed circumnuclear disc}
\label{sec:origin_broad_abs}

The absorption feature centred at the systemic velocity has a high FWZI $= 360$ \kmps, and is detected across the entire radio source (see Figs. \ref{fig:wsrt_evn_int} and \ref{fig:vel_disp_outflow}). These together imply that the gas causing this absorption is close to the radio source and screens it in a smooth distribution at the resolution of our observations (\apx 4 pc). However, the shallow blueshifted absorption wing, which is indicative of outflowing gas, is detected only towards a small part of the eastern lobe, suggesting the presence of clumps of gas as well. 

HST observations by \citet{Perlman01} show the presence of a dust lane almost perpendicular to the radio axis (i.e. PA $\sim 200^\circ$; see Fig. \ref{fig:Optical}). Molecular gas observations further suggest the presence of a warped gas disc that is almost edge-on with a position angle of 5$^{\circ}$ -- 10$^{\circ}$, at \apx 1.4 kpc \citep{Garcia-Burillo07}, changing to PA$=350^\circ$ at \apx 10 kpc \citep[][ Murthy et al. in preparation]{Morganti23a}. Thus, it is likely that the \hi\ too belongs to this circumnuclear gas disc.

\citet{Conway96} and \citet{Struve12} previously proposed this scenario based on the difference in \hi\ column density along the two lobes, and the detection of a tentative velocity gradient in \hi\ in their VLBA data along the eastern lobe at PA$=100^\circ$, which is almost along the radio axis. \citet{Struve12} further estimated a scale height of \apx 50 pc for this disc.  

With more sensitive observations, however, we find that while the velocity gradient does indeed exist, it is along an axis perpendicular to the radio axis. This can be seen in Fig. \ref{fig:vel_field}, the velocity field of \hi. The velocity gradient is present along the western lobe and parts of the eastern lobe close to the core at PA \apx\ $5^\circ - 10^\circ$, which shows similarities to the kpc-scale disc reported by \citet{Garcia-Burillo07}. However, this gradient is absent along the rest of the eastern lobe, including the radio hot spot, suggesting that the regular rotation of the gas is significantly disrupted there due to the eastern lobe interacting with the gas disc. Furthermore, the velocity dispersion is high --- at $\geq40$~\kmps\ , as compared to a typical value of 20 \kmps\ in the central regions of early-type galaxies --- for most of the gas along both lobes, implying that the entirety of the circumnuclear gas is highly turbulent.

Based on our findings, we suggest that the radio source is expanding into the circumnuclear disc. The eastern lobe is interacting directly with one or more gas clouds, disrupting the regular rotation of the gas locally. The western lobe, although diffuse in morphology, consists of a weaker hot spot, indicating the presence of a milder interaction with the gas. Overall, the gas has a high velocity dispersion everywhere ($\geq 40$~\kmps; Fig. \ref{fig:vel_disp_outflow}, left), suggesting that the jet--ISM interaction has affected the disc significantly. Furthermore, the asymmetric morphology of the radio source could be due to the two radio lobes encountering different environments in the process of expansion. The direct jet--cloud interaction occurring along the eastern lobe is also driving an outflow, which we discuss in the following subsection.

\subsection{Parsec-scale \hi\ outflow}
\label{sec:outflow}

Given the strong jet--ISM interaction taking place along the eastern lobe, it is very likely that the shallow blueshifted absorption wing arises from an outflowing gas cloud. In the WSRT spectrum, this feature extends up to \apx 400 \kmps\ , while at VLBI scales we only recover the feature extending up to \apx 300 \kmps, and the signal is also weaker. As the outflow is resolved even at parsec scales (see Fig. \ref{fig:vel_disp_outflow}), it is  likely that the reduced optical depth sensitivity of our VLBI observations is the cause of this partial detection.

This outflow is detected only in a small region along the eastern lobe \apx 35 pc from the core  and is not detected against the radio hot spot where the optical depth sensitivity is the highest (Fig.~\ref{fig:vel_disp_outflow}). This suggests that the outflowing gas is clumpy. Similar clumpy gas outflows at parsec scales have been reported in other young or restarted radio sources \citep[e.g.][]{Morganti13, Schulz18, Schulz21}. 

Following \citet{Mahony13} and \citet{Heckman02}, we estimate a mass-outflow rate ($\dot{M}$) using the expression:
\begin{equation}
   \dot{M} = 30 \frac{\Omega}{4\pi} \times \frac{r_*}{1kpc} \times \frac{N_{\hi}}{10^{21} \rm cm^{-2}} \times \frac{v}{300 \rm km s^{-1}}~\rm M_\odot~year^{-1}
,\end{equation}
where $\Omega$ is the solid angle that the outflow subtends which we assume to be equal to $\pi$, $r_*$ is the radius of the outflow which we measure from the core to be 35 pc, \NHI\ is the \hi\ column density, and $v$ is the velocity of the outflow. 

We use the Gaussian components of the VLBI absorption spectrum shown in the bottom-right panel of Fig.~\ref{fig:vel_disp_outflow}, from which we obtain an FWHM for the outflow of 134 \kmps\ with the velocity centroid being $-300$ \kmps. We obtain a peak absorption of 1.0 mJy beam\p{1} from the fit. The total flux from the region where the outflow is seen is 210 mJy. We estimate an integrated optical depth for the outflow using the approximation: $\int \tau dv = 1.06 \times FWHM \times \tau_{peak} = 0.68$ \kmps. Closer to the AGN and with disturbed kinematics, as in this case, the spin temperature of the gas is generally higher than the typically assumed 100K \citep[see][]{Morganti18}. Thus, assuming \tspin $= 1000$ K for the outflow gives a total \hi\ column density of \NHI $= 1.2 \times 10^{21}$ cm\p{2}. This gives us a mass-outflow rate of \apx 0.3 \Msun\ year\p{1}. As we do not detect the entire outflow, our estimate of $\dot{M}$ is only a lower limit. If we assume that the entire blueshifted wing seen in the WSRT spectrum arises from the outflowing gas in the same region then we obtain an upper limit to the mass-outflow rate. Fitting Gaussian components to the WSRT spectrum gives a peak absorption of $4.0$ mJy beam\p{1}, FWHM $= 218$ \kmps\ , and a velocity centroid of $-205$ \kmps\ for the blueshifted wing. Using these, and assuming \tspin $= 1000$ K, we obtain $\dot{M} = 1.4$~\Msun~year\p{1}. \citet{Ocana-Flaquer10} report a star formation rate (SFR) of 4.9 \Msun\ yr\p{1} based on far-infrared measurements. Thus if the \hi\ mass-outflow rate is close to the derived upper limit of 1.4 \Msun\ yr\p{1}, it could have an impact on quenching the star formation in the galaxy. However, the reported SFR is an average over the entire galaxy. Our observations, in comparison, cover only a small region of the centre of the galaxy. Thus, given the different scales, a fair comparison between the two quantities is not possible. We can also obtain an order-of-magnitude estimate of the mass of the gas constituting the outflow, assuming a spherical geometry for the gas cloud. The radius of the gas cloud measured based on the size of the region where the outflow is detected is \apx 10pc giving us a mass of \apx $10^4$ \Msun. The total gas mass in the circumnuclear region of this galaxy is \apx\ $6 \times 10^8$ \Msun\ \citep{Ocana-Flaquer10}. Thus, the outflow does not constitute the bulk of the gas in the circumnuclear region in this case. This has been shown to be case in other sources with pc-scale VLBI \hi\ absorption studies (e.g. 4C\,12.50, 3C\,236), where the detected gas clouds have masses of \apx $10^5$ \Msun\ and the pc-scale outflows (at most tens of \Msun\ year\p{1}) constitute only a small fraction of the circumnuclear gas \citep[e.g.][]{Morganti13, Schulz21}.

\subsection{Jet-induced feedback}

\label{sec:jet-ism_interaction}

 \hi\ outflows have only been studied from kpc to pc scales in four sources so far, namely 4C 12.50, 3C 236, 4C 52.37, and 3C 293 \citep[][]{Morganti13, Schulz18, Schulz21}. The estimated ages of these sources range between \apx $10^4$ and $10^5$ years, and in all these sources, the \hi\  is found to be clumpy at pc scales. In the younger sources, the entire outflow is concentrated in the central few pc, while in relatively older sources, only part of the outflow is recovered at pc scales, with the rest of the outflow being spread out over hundreds of pc, resolved out at VLBI scales. \target\ is at least an order of magnitude younger than these sources, where --- while gas clumps are indeed present --- most of the \hi\ is in a continuous distribution even at a resolution of 4 pc. In all these cases, the pc-scale outflows do not appear to be the dominant form of negative feedback from radio jets. Instead, the main impact of the radio jets on the ambient ISM is the increased turbulence of the gas present in the circumnuclear region.

These findings, albeit drawn from only a handful of sources, agree with numerical simulations of jets expanding into gas discs, which predict an evolution in the nature of the jet--ISM interaction over the age of a radio source \citep[e.g.][]{Wagner11, Wagner12, Mukherjee18b, Dutta24}. Moreover, \citet{Sutherland07} claim that the radio morphology of \target\ matches that of a radio source in one of the earliest phases of evolution in their simulations. These authors predict that radio jets in this case are interacting strongly with an asymmetric ISM, creating an expanding bubble of plasma and gas. The western jet is about to break out of this bubble after interacting with the ISM, while the interaction is still strong along the eastern lobe where the radio jet is still within the created bubble. 

Our results strongly support this scenario. We find that in \target\ the pc-scale radio lobes are expanding into circumnuclear gas and are interacting directly with an asymmetric but predominantly smoothly distributed ISM along the two lobes. Along the eastern lobe, the jet has encountered gas clouds and the resulting interaction has disturbed the gas and is also driving a cold gas outflow at \apx 35 pc from the core, in projection. The optical spectrum of \target\ exhibits only a few weak emission lines \citep{Marcha96}. Moreover, \citet{Zovaro19} found that the line ratios of the near-infrared lines are consistent with shock excitation rather than photoionisation. Thus, the effect of radiation on the ambient gas is not likely to be significant, making the radio jets the main cause of the disturbed kinematics of the gas. Also, we find no strong jet--ISM interaction along the western lobe, although it has a hot spot and the gas in this region is highly turbulent ($\sigma \geq 40$ \kmps), suggestive of a strong interaction in the past. 

These simulations also predict that the radio plasma that percolates through the disc during the expansion of the jets drives shocks into the ISM over spatial scales that are larger than the apparent extent of the radio source. Although \hi\ absorption observations are very useful for studying the kinematics of the gas at very high spatial resolution, the spatial extent of the gas that can be studied is limited to the size of the background radio source. Thus, to understand the impact of the radio jets at larger spatial scales, this needs to be combined with complementary emission studies that do not suffer from the same limitation. In the case of \target, warm molecular gas observations by \citet{Zovaro19} have already shown that the gas exhibits clear signatures of shock excitation even at kpc scales owing to the impact of the radio jets, which is in agreement with the aforementioned numerical simulations.

\subsection{Survival of atomic gas in subkpc-scale outflows}

Although these simulations are able to reproduce the observed kinematics of the circumnuclear gas remarkably well \citep[e.g.][and references therin]{Mukherjee18a, Mukherjee18b, Zovaro19, Murthy22b}, they only consider gas at very high temperatures ($>10^5$ K), while the atomic and molecular gas that is a part of the ISM directly interacting with the radio jets is expected to be at lower temperatures \citep[$<10^5$ K; e.g.][]{Perucho24}. Thus, it still needs to be verified whether the conditions defined in these simulations are conducive for the survival of cold gas and, as such, are a good representation of the actual situation. 

\citet{Perucho24} provide an analytical solution to this difficult problem with a focus on atomic gas outflows. These authors predict that the shocks driven by low-power radio jets ($\leq 10^{43}$ erg s\p{1}) are not strong enough to ionise entire atomic gas clouds at any spatial scale at any given time. Moreover, the cooling time for the ionised gas in the clouds is almost instantaneous compared to the lifetime of the source. Thus, atomic gas should exist in outflows even within the central subkpc region affected by the expanding radio jets either via direct interaction or shocks. 

On the other hand, the cold gas may or may not be detected at subkpc scales in higher power sources ($>10^{43}$ erg s\p{1}) depending on the density of the gas. Such sources are capable of ionising a large fraction of the gas and also cause the cooling times to be longer due to the mixing of the clouds with the hotter environment. At low gas densities ($<10^3$ cm\p{3}), this would result in the atomic gas outflows being detected only at kpc scales. However, when the density of the gas is higher ($>10^3$ cm\p{3}), the post-shock temperature is lower and the cooling times are very short compared to the lifetime of the source (see Fig.4 in \citet{Perucho24}). This results in rapid recombination allowing the atomic gas to exist even in the extreme conditions at subkpc scales.

We estimate the jet power of \target\ to be in the range of $6 \times 10^{43}$ erg s\p{1} to $10^{44}$ erg s\p{1} using the relations presented in \citet{Cavagnolo10} and \citet{Wu09}. As these estimations could at best be a lower limit \citep[e.g.][]{Mukherjee18b}, we note that \target\ is an intermediate or high-power radio source. Thus, in the scenario presented by \citet{Perucho24} the presence of a subpc outflow in \target\ indicates that the gas is sufficiently dense for the cooling times to be very short and the recombination rate to be high, while the mixing between the cold and hotter gas via instabilities is also likely to be inefficient.

Thus, overall, these results add to the growing body of evidence that radio jets have a significant impact on cold gas in the early stages of their evolution. In the very central region, radio jets primarily make the gas highly turbulent, while also driving outflows, albeit only constituting a small fraction of the total circumnuclear gas. Furthermore, the overall impact on the gas spans a much larger spatial scale --- even an order of magnitude larger in the case of \target\ --- than the apparent extent of the radio source itself. Similar studies of radio sources spanning a wide range in age, radio luminosity, and size are needed to build a clearer picture of the nature of the jet--ISM interaction and the impact it has on the host galaxies.

\section{Summary}

We present deep \hi\ absorption observations at kpc and pc scales using the WSRT and the global VLBI array of a young radio AGN, \target, a well-known CSO with radio lobes extending up to \apx 140 pc. At kpc scales, we detect a narrow absorption (FWZI =$6.6$ \kmps) redshifted by 220 \kmps\ with respect to the systemic velocity estimated using $z=0.0598$, a broad absorption centred at the systemic velocity, and a shallow blueshifted wing with a combined FWZI of 525 \kmps. This study is focused on the broad absorption at the systemic velocity and the blueshifted wing, as these correspond to the gas in close proximity to the radio lobes. We measure a radio continuum flux density of $(2.67 \pm 0.267)$ Jy and a total \hi\ column density of the gas --- excluding that giving rise to the narrow absorption profile --- of $(1.04 \pm 0.09) \times$ 10\pp{21} cm\p{2}. 

At pc scales with our Global VLBI observations, we detect the redshifted absorption and the broad absorption at the systemic velocity. We measure a total continuum flux density of $(1.19 \pm 0.2)$ Jy with our VLBI data. The absorption at the systemic velocity is well resolved at the resolution of our VLBI observation and detected across the entire radio source. We also detect a part of the blueshifted wing but only against a small region of the eastern lobe, \apx 35 pc from the core, just prior to the radio hot spot. The gas has high velocity dispersion everywhere ($\geq 40$ \kmps) and the region of the highest velocity dispersion (\apx 90 \kmps) includes the location of the outflow and the region containing the radio hot spot along the eastern lobe. We detect a velocity gradient along the western lobe and parts of the eastern lobe close to the core at PA \apx\ $ 5^\circ - 10^\circ$, while such a gradient is considerably weaker along the rest of the eastern lobe. We also find that the \hi\ column density is higher along the eastern lobe, with an average \NHI $= 2.5 \times 10^{21}$ cm\p{2} compared to the western lobe where the average \NHI\ is 7 $\times$ 10\pp{20} cm\p{2}. 

Combining the above findings, we arrive at the conclusion that the lobes of \target\ are expanding into a circumnuclear disc, partially disrupting it along the eastern lobe, and overall, making the gas highly turbulent. While the gas is predominantly in a smooth distribution at the resolution of our data (\apx 4 pc), being detected all across the radio source, we also find the presence of clumps at least along the eastern side. Here, the eastern lobe is strongly interacting with gas clouds and also driving an outflow of an \hi\ cloud. We estimate a mass-outflow rate of $0.3~\leq~\dot{M}~\leq~1.4$~\Msun~year\p{1}. We detect only a part of the outflow compared to that detected by the WSRT. We suggest that this is due to lower optical depth sensitivity of our observations on account of the outflow being resolved. 

Our results support the scenario proposed by \citet{Sutherland07} where \target\ is in a very early stage of expansion into an asymmetric ISM. More broadly, our results also agree with the predictions of similar theoretical models regarding the nature of the ISM and the nature of jet--ISM interaction in young radio sources expanding into the circumnuclear gas. 

Furthermore, we compare our observations with the predictions of a recent analytical model of the impact of radio jets on atomic gas by \citet{Perucho24}. We find that the existence of a subkpc-scale outflow in this case could be due to inefficient mixing of the cold gas with the hot medium, and high gas density leading to an almost instantaneous recombination of the ionised gas.

Overall, our study demonstrates the importance of probing the jet--ISM interaction at multiple spatial scales and supports the findings presented of earlier studies that radio AGN in the early stages of their evolution significantly impact the gas into which they are expanding, especially that in cold phase.

\begin{acknowledgements}

We thank the referee for insightful comments that improved the presentation of the paper. We also thank Manel Perucho for an interesting and useful discussion on sub-kpc \hi\ outflows. SM thanks the organisers of the Lorentz Centre workshop, `The importance of jet-induced feedback on galaxy scales' where interesting discussions regarding some of the results presented in this paper were held. The European VLBI Network is a joint facility of independent European, African, Asian, and North American radio astronomy institutes. Scientific results from data presented in this publication are derived from the following EVN project code: GS045. The National Radio Astronomy Observatory is a facility of the National Science Foundation operated under cooperative agreement by Associated Universities, Inc. The Westerbork Synthesis Radio Telescope was operated by ASTRON (Netherlands Institute for Radio Astronomy) with support from the Netherlands Foundation for Scientiﬁc Research (NWO).

\end{acknowledgements}

%
   \bibliographystyle{aa} 
   \bibliography{ref} 
%


\end{document}